\documentclass{emulateapj}
\usepackage{graphicx}
\usepackage{amssymb}
\usepackage{amsmath}
\usepackage{ulem}
\usepackage[colorlinks=true,linkcolor=blue,anchorcolor=blue,citecolor=blue,urlcolor=blue]{hyperref}  

\shorttitle{Strongly Lensed Jets, Time Delays, and the Value of $H_0$}
\shortauthors{Barnacka et al.}

\begin{document}

\title{Strongly Lensed Jets, Time Delays, and the Value of H$_0$}

\author{Anna Barnacka$^{1,2}$}
\author{Margaret J. Geller$^1$}
\author{Ian P. Dell'Antonio$^3$}
\author{Wystan Benbow$^1$}
\affil{$^1$Harvard-Smithsonian Center for Astrophysics, 60 Garden St, MS-20, Cambridge, MA 02138, USA\\
$^2$Astronomical Observatory, Jagiellonian University, Cracow, Poland \\
$^3$Department of Physics, Brown University, Box 1843, Providence, RI 02912}

\email{abarnacka@cfa.harvard.edu}

\begin{abstract}

In principle, the most straightforward method of estimating the Hubble constant relies on  time delays 
between mirage images of strongly-lensed sources.  
It is a puzzle, then, that  the values of  H$_0$ obtained with this method span a range 
from $\sim 50 - 100$~$\mbox{km\,s}^{-1}$Mpc$^{-1}$. 
Quasars monitored to measure these time delays, are multi-component objects. 
The variability may arise from different components of the quasar or 
may even originate from a jet. 
Misidentifying a variable emitting region in a jet with emission from the core
region 
may introduce an error in the Hubble constant derived from a time delay.
Here, we investigate the complex structure of  sources as the underlying physical explanation 
of  the wide spread  in values of the Hubble constant based on gravitational lensing.
Our Monte Carlo simulations demonstrate that
the derived value of the Hubble constant is very sensitive to the offset between the center of the emission 
and the center of the variable emitting region.
Thus, we propose using the value of H$_0$ known from other techniques to spatially resolve 
the origin of the variable emission once the time delay is measured. 
We advocate  this method particularly for gamma-ray astronomy, 
where the angular resolution of detectors reaches approximately 0.1~degree;
lensed blazars offer the only route  for identify the origin of gamma-ray flares. 
Large future samples of  gravitationally lensed sources identified with Euclid, SKA, and LSST 
will enable a statistical determination of H$_0$. 

\end{abstract}


\section{Introduction}

The Hubble constant, H$_0$,
is a fundamental cosmological parameter. 
A rich variety of observational methods yield remarkably precise values of this parameter \citep{2010ARA&A..48..673F}.
For example, 
the Cepheid distance ladder gives  $73.8\pm$2.4~$\mbox{km\,s}^{-1}$Mpc$^{-1}$ \citep{2011ApJ...730..119R,2011ApJ...732..129R}.

In principle, gravitational lensing provides an independent one-step method for Hubble constant determination.
This approach was first proposed by \citet{1964MNRAS.128..307R}, 
who suggested using  time delays between the images of gravitationally-lensed supernovae 
 long before discovery of the first gravitationally lensed quasar \citep{1979Natur.279..381W}.
The time delays are proportional to H$_0^{-1}$ and weakly depend on other cosmological parameters 
\citep{1964MNRAS.128..307R,1997ApJ...475L..85S,2002MNRAS.337L...6T,2002ApJ...578...25K,2003ApJ...599...70K,2007ApJ...660....1O,2013ApJ...766...70S,2014MNRAS.437..600S}.

Initially, the practical implementation of the time delay method suffered from a number of challenges 
\citep{2003astro.ph..4497C,2004mmu..symp..117K,2005IAUS..225..281S}. 
The accuracy of the method relies on the precision of the  time delay determination, 
knowledge of the redshifts in the system and  reconstruction of the mass distribution of the lens.
Long-term monitoring of gravitationally-lensed quasars
provides  time delays with uncertainties of a few percent  
\citep{2002ApJ...581..823F,2011A&A...536A..44E,2013A&A...557A..44R,2013A&A...556A..22T,2013A&A...553A.121E}.
Advances in spectroscopy combined with  precise  cosmology allow for distance measurement 
with great accuracy.
The mass distribution of the lens can be  reconstructed well using resolved radio and optical images.
In addition, for lenses located at a low enough redshift, 
the velocity dispersion of the lens can be measured,
which allows for independent confirmation of the mass distribution \citep{2008ApJ...684..248B}.

To date, dozens of gravitationally-lensed systems yield measured time delays. 
Lens modeling of these systems provides estimates of the Hubble constant 
covering a large range from $\sim 50$ to $\sim 100$~$\mbox{km\,s}^{-1}$Mpc$^{-1}$ 
\citep{2014arXiv1404.2920R,2010ApJ...712.1378P,1999ApJ...527..498F}. 
It is puzzling that these estimates of the Hubble parameter derived from time delays span a range much larger than expected from other precise astrophysical methods.

Measurements of time delays are based on  monitoring  variable sources.
The variability of quasars has been known for a long time \citep{1963ApJ...138...30M}.
The variability of quasars can be described well by a damped random walk \citep{2010ApJ...721.1014M,2010ApJ...708..927K}
and the optical flux fluctuations are interpreted as thermal fluctuations driven by an underlying stochastic process \citep{2009ApJ...698..895K}.

The amplitude of variability in  radio-quiet quasars is usually small. 
 \citet{2007AJ....134.2236S} reported that
at least 90\% of quasars are variable at the 0.03 mag level (rms) on
timescales up to several years,
and that 30\% of quasars are variable at the 0.1 mag level.
The variability of gravitationally-lensed quasars  selected for monitoring by  
COSMOGRAL\footnote{the COSmological MOnitoring of GRAvItational Lenses: http://cosmograil.epfl.ch/} is typically  in the range from 0.2 to even 2 mag.

Monitoring of lensed systems  focuses on radio-loud quasars,
which constitute about 20\% of the quasar sample.  
The emission of radio-loud quasars is  not limited to thermal emission from the accretion disk;
the majority of the variable emission may originate from  relativistic jets.

These jets are the largest particle accelerators in the universe,
producing radiation from radio up to very high energy gamma-rays.
Observations of jets  from radio to x-ray wavelengths, where the sources are resolved, 
reveal that the jets have very complex structure composed of bright knots, blobs and filaments,
with sizes ranging from the subparsecs up to dozens of kpcs
\citep{2006ARA&A..44..463H,2007ApJ...662..900T,2002ApJ...570..543S,2008Natur.452..966M,1991AJ....101.1632B}.  
The source variability can be very complex. 
Variable emission may originate from different components close to the core, 
or it can originate from the knots along the jet as in the well studied example of M87 \citep{2006ApJ...640..211H}.

If the projected distance between the central part of a quasar and the variable emission sites along a jet  
are even  1$\%$ of the Einstein radius, which corresponds typically to $\sim30\,$pc, 
 time delays and magnification ratios between 
the images may differ significantly. 
These differences, in principle, can be used to investigate the structures of  jets at high energies,
where the emission cannot be spatially resolved due to 
the poor angular resolutions of the detectors \citep{2014arXiv1404.4422B,2014arXiv1403.5316B}.
Conversely, the multiple emitting regions of quasars
 can affect the determination of the Hubble constant.
 
Here, we investigate the impact of the complex structure of the source on Hubble parameter determination. 
We summarize the current  Hubble parameter measurement in \S~\ref{sec:H0}.
To demonstrate how  estimates of the Hubble constant vary with the projected distance between the core and 
the emitting region within the jet, we  build a toy model of the M87 jet, which we describe in \S~\ref{sec:TM}.
We  give an overview of strong gravitational lensing phenomena in \S~\ref{sec:SL}, 
and perform Monte Carlo simulation in \S~\ref{sec:MC}. 
We discuss the results of simulations in \S~\ref{sec:dis}.
We summarize  existing lensing measurements of H$_0$ (\S~\ref{sec:dis:H0}).
Then, in \S~\ref{sec:dis:tuneH0}, we propose to use a Hubble parameter tuning method to elucidate 
the spatial origin of variable emission. 
We discuss the application of the Hubble parameter tuning approach to high-energy astronomy in \S~\ref{sec:dis:gamma}.
We conclude in \S~\ref{sec:con}.

\section{Hubble Constant}
\label{sec:H0}

\begin{table}
\caption{Hubble constant measurements  for individual gravitationally lensed sources.}
\label{tab:H0}
\scriptsize
\begin{center}
\begin{tabular}{lcl}
  \hline                       
 Object  & Hubble constant  & References \\
  \hline \hline
HE 0435-1223       & $91.9\pm17.1$ & \citet{2014arXiv1404.2920R} \\
                                 & $62 \pm 5$ & \citet{Courbin2011} \\
RX J0911.4+0551 & $79.3\pm32.6$ & \citet{2014arXiv1404.2920R} \\
                                  & $71\pm12$ & \citet{2002ApJ...572L..11H} \\   
SBS 0909+532      & $81.9\pm20$ & \citet{2013ApJ...770..154T} \\     
FBQ 0951+2635    & $60\pm9$ & \citet{Jakobsson2005} \\
Q0957+561            & $50\pm17$ & \citet{1991Natur.350..211R} \\
                                  & $77\pm29$ & \citet{1999AJ....118...14B} \\                                     
                                  & $79.3\pm7$ & \citet{2010ApJ...711..246F} \\       
                                   & $97.3\pm31.4$ & \citet{2014arXiv1404.2920R} \\     
SDSS J1004+4112  & $91.8\pm30.4$ & \citet{2014arXiv1404.2920R} \\          
HE 1104-1805          & $62\pm4$ & \citet{Merino2002} \\         
PG 1115+080           & $59\pm12$ & \citet{2002MNRAS.337L...6T} \\ 
                                    & $57\pm7$ & \citet{2004MNRAS.354..343T} \\     
                                    & $61.5\pm19.5$ & \citet{2014arXiv1404.2920R} \\   
RX J1131-1231       & $71.6\pm25.4$ & \citet{2014arXiv1404.2920R} \\
			       & $78.7^{+4.3}_{-4.5}$ & \citet{2013ApJ...766...70S} \\	
SBS 1520+530        & $58.3\pm17.3$ & \citet{2014arXiv1404.2920R} \\
B1600+434               & $74\pm14$ & \citet{Koopmans2000} \\
B1608+656              & $63\pm15$ & \citet{2002ApJ...581..823F} \\
                                   & $75\pm6$ & \citet{2003ApJ...599...70K} \\
                                   & $69.7\pm5$ & \citet{2010ApJ...711..201S} \\
                                   & $60.3\pm11.2$ & \citet{2014arXiv1404.2920R} \\
SDSS J1650+4251& $51\pm7$ & \citet{Vuissoz2007} \\
WFI J2033-4723     & $63\pm5$ & \citet{Vuissoz2008} \\
                                   & $71.5\pm12.2$ & \citet{2014arXiv1404.2920R} \\
B 0218+357             & $69\pm15$ & \citet{1999MNRAS.304..349B} \\
                                   & $76\pm7$ & \citet{2000ApJ...536..584L} \\
                                   & $78\pm6$ & \citet{2004MNRAS.349...14W} \\
                                   & $61\pm7$ & \citet{2005MNRAS.357..124Y} \\
                                   & $64\pm4$ & \citet{2014ApJ...782L..14C} \\
PKS 1830+211       & $76\pm18$ & \citet{1999ApJ...514L..57L} \\
                                   & $73\pm35$ & \citet{2000ApJ...536..584L} \\
                                   & $44\pm9$ & \citet{2002ApJ...575..103W} \\
                                   & $48.7\pm13.7$ & \citet{2013arXiv1307.4050B} \\
HE 2149-2745        & $86.8\pm33.5$ & \citet{2014arXiv1404.2920R} \\
                                   & $65\pm8$ & \citet{Burud2002} \\
  \hline  \hline

\end{tabular}

\end{center}
\end{table}
\normalsize

The current era of high-precision cosmology provides measurements 
of the Hubble constant, H$_0$, with a remarkably small error. 
For example, 
 the modeling of Planck observations of the cosmic microwave background fluctuations using the flat-$\Lambda$CDM cosmological model 
gives H$_0=67.3\pm1.2\,$~km$\,$s$^{-1}\,$Mpc$^{-1}$ \citep{2013arXiv1303.5076P}.

Many independent methods provide a measure of H$_0$. 
Examples include 
the Hubble Space Telescope Key Project which provided a H$_0=72\pm$8~$\mbox{km\,s}^{-1}$Mpc$^{-1}$ \citep{2001ApJ...553...47F}. 
The Cepheid distance ladder gives  $73.8\pm$2.4~$\mbox{km\,s}^{-1}$Mpc$^{-1}$ \citep{2011ApJ...730..119R,2011ApJ...732..129R}
and $74.3\pm1.5\,(\mbox{statistical})\pm 2.1\,(\mbox{systematic})\, \mbox{km\,s}^{-1}$ Mpc$^{-1}$ \citep{2012ApJ...758...24F}.

One of the most straightforward methods of estimating the value of H$_0$ relies on  time delays 
between images of strongly-lensed sources. 
This method has the advantage of
being independent of the local distance ladder.
However, individual lens systems  yield varied results (see Table~\ref{tab:H0}). 
\citet{2010ApJ...711..246F} plot the H$_0$ obtained for the well-observed set of individual lens systems. 
Roughly half of the  studies are consistent with 
H$_0 < 60$ ~$\mbox{km\,s}^{-1}$Mpc$^{-1}$, well outside the limits from other measures.  
The rest scatter across the range H$_0 = 65-80$~$\mbox{km\,s}^{-1}$Mpc$^{-1}$. 
Thus these measurements present a puzzle and a challenge to understand the astrophysics that might underlie the varying results.

Studies of  Q0957+561 \citep{1999AJ....118...14B,2000ApJ...542...74K}
 result in  H$_0 > \,$85~$\mbox{km\,s}^{-1}$Mpc$^{-1}$,
 and studies  of PKS 1830-211  result in values across the range H$_0 = 44-76$~$\mbox{km\,s}^{-1}$Mpc$^{-1}$\citep{1999ApJ...514L..57L,2000ApJ...536..584L,2002ApJ...575..103W,2013arXiv1307.4050B}.
Both of these systems host powerful jets.
More recently, \citet{2010ApJ...711..246F}  analyzed  HST data for Q0957+561 and
identified 24 new strongly lensed features,
which they use to constrain a mass model for the lens. 
Adopting the radio time delay measured with an error $\lesssim 1$\%, 
they found H$_0=79.3^{+6.7}_{-8.5}\mbox{km\,s}^{-1}$Mpc$^{-1}$.
\citet{2010ApJ...711..246F} then concluded that the quasar flux ratio predicted by their
detailed lens model 
is inconsistent with existing radio measurements again indicating that there is some poorly understood underlying astrophysics.

Another system, B0218+357,  "a golden lens" for  H$_0$ measurement \citep{2004MNRAS.349...14W},
also hosts a powerful jet.
The H$_0$  values based on this systems are in the range 61-78~$\mbox{km\,s}^{-1}$Mpc$^{-1}$ 
\citep{2005MNRAS.357..124Y,2014ApJ...782L..14C,2000ApJ...536..584L,2004MNRAS.349...14W}.
The most recent attempt to measure H$_0$ for this system,  using the time delay of $11.46\pm0.16\,$days from gamma-ray emission,
results in a Hubble constant of $64\pm4$~$\mbox{km\,s}^{-1}$Mpc$^{-1}$ \citep{2014ApJ...782L..14C}.

\citet{2013A&A...559A..37S,2014A&A...564A.103S} investigated the impact of mass-sheet
degeneracies on the predictions of the Hubble constant and found that
these mass-modeling errors could lead to deviations of up to 20\% on
the Hubble constant.
\citet{2013arXiv1306.4732S} investigate the tension between the H$_0$ obtained using gravitational lensing and the Planck results.
They revise analysis of the gravitational lens RXJ1131-1231, 
reducing the systematic errors introduced by an assumed lens model density profile.
They emphasize  that the tension in H$_0$ measurements from different methods 
could be due to unknown systematic uncertainties.

The observational situation presents a mystery. 
The straightforward application of lensing time delays produces an array of results for 
the Hubble constant that are often substantially offset from those obtained with other standard methods. 
The scatter among the lensing results is also remarkably large. 
Previous attempts to explore the scatter in the values of H$_0$ from
individual lenses have assumed sources that vary uniformly, 
so that the centroid of the light distribution is also the centroid of the variability.
We investigate the structure of the sources as an explanation for these puzzling results.

\section{Toy Model: Lensed Jet}
\label{sec:TM}

\begin{figure}
\begin{center}
\includegraphics[width=9cm,angle=0]{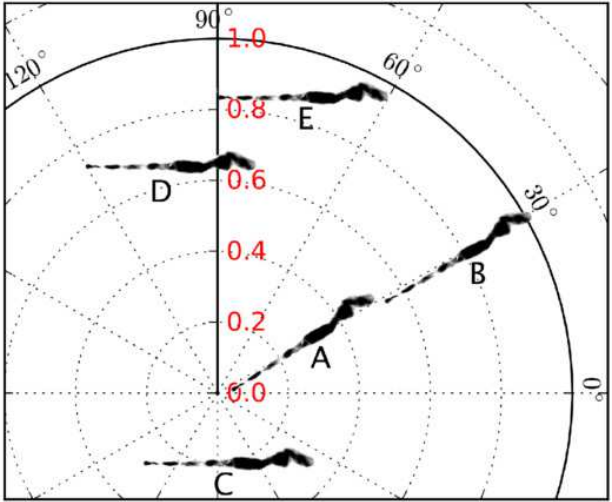}
\end{center}
\caption{\label{fig:image_planeSIS} 
                           Example alignments of  jets in the source plane.
                           The radial coordinates indicate the distance from the  center of the lens mass
                           normalized by the Einstein radius. }
\end{figure}

Here we introduce a toy model of a strongly lensed source 
to investigate  the impact of complex source structure 
on  H$_0$ estimation.
As an inspiration for the source model, we  use the nearest and best studied 
active galactic nucleus with a relativistic jet, M87. 

M87 is a local example of a source with very complex variability \citep{2006Sci...314.1424A,2012ApJ...746..151A}.
Spatially-resolved  observations of M87 reveal that variable emission 
may originate from at least two locations: the core and the HST-1 knot embedded within the jet. 
The core refers to the central region of the active galactic nucleus. 
Commonly the term "core" is defined as the position of the brightest feature 
in VLBA images of blazars \citep{1998A&A...330...79L,2008ASPC..386..437M,2010ApJ...722L...7P}.
The core and HST-1 have a projected separation of $\sim$60~pc \citep{1999ApJ...520..621B}.
The M87 jet also has many bright  knots at much greater distances than HST-1. 

During the flaring activity of M87 observed by the Chandra satellite in 2004/2005,
the x-ray flux from HST-1 increased by a factor of 50 \citep{2006ApJ...640..211H};  
thus demonstrating that the variability can originate from different parts of the jet. 
The locations of the variable emission can be
 hundreds of parsecs away from the central engine. 

M87 is located at 16~Mpc \citep{1991ApJ...373L...1T}; 
thus the separation between the emitting regions of 60~pc corresponds to 0.7 arcseconds.  
The angular resolution of the Chandra satellite is 
0.5$\,$arcseconds\footnote{http://cxc.cfa.harvard.edu/cdo/about\_chandra/overview\_cxo.html},   
allowing discrimination between emission from HST-1 and from the core. 
If we imagine an M87 analog located at 1~Gpc from the observer, 
this separation would correspond to  0.01 arcseconds.
The emission would  be unresolved in the x-rays and it would be challenging to resolve even with HST. 

Lensed sources are usually located at distances greater than 1~Gpc. 
The M87 example demonstrates that variability originating from  regions along the jet could easily be misinterpreted 
as emission from the core.

The angular separation between the core and knots along the jet
depends on the viewing angle to the jet and the distance of the knot from the core.   
The emission from the jet is relativistically boosted when 
the jet is pointed at a small angle ($\lesssim$ 20$^{\circ}$ - 30$^{\circ}$) 
relative to the line of sight between the source and the observer.
The viewing angle for a typical radio-loud quasars is $<5^\circ$,
for BL Lacs objects it is  $<10^\circ$,
and it is $<50^\circ$ for the radio galaxies \citep{1999ApJ...521..493L}.
The viewing angles for a sample of the brightest {\it Fermi}/LAT detected 
blazars derived from high-resolution VLBA images range from 1$^{\circ}$ to 5$^{\circ}$ \citep{2010A&A...512A..24S}. 
The M87 jet is observed at $\sim15^\circ$ relative to the line of sight \citep{1999ApJ...520..621B,2011ApJ...743..119P};
it is thus representative of the range of observed sources.

Luminous knots may be located even hundreds of kpcs  from the central engine \citep{2011ApJS..197...24M}.
Thus, even when the viewing angle is small, of the order of a few degrees, 
the angular separation between distant knots along the jet  may constitute a significant fraction of the Einstein radius of the lens, 
or may extend beyond it.

The geometry of the sources can be very complex.
There are also many different configurations of the source-lens-observer systems.
However, in the source plane, the parameter space that adequately describes these sources 
reduces to the separation between the core and the variable knot.
In our toy model we randomly select both the position of the core in the source plane, 
and the alignment of jet relative to the core.
We investigate the resulting changes in the time delay and the magnification ratios.

To investigate the impact of source structure on  the determination of H$_0$  further,
we consider an M87-like source, located at redshift $z_S=2.5$, 
and a lens at redshift $z_L=0.89$. 
Our choice of lensing system is motivated by the observed case of the gravitationally-lensed blazar PKS~1830-211,
where gamma-ray emission was used for the first time to investigate gravitational lensing at high energies \citep{2011A&A...528L...3B}.
 We consider what happens to the value of H$_0$ as we vary the projected distance between the core and 
the variable emitting region along the jet. 
We set the mass of the lens so that the Einstein radius for a source at z=2.5 is 0.4".

The measured quantities include the distance ratio, the time delay, the magnification ratio, and the morphology of the lensed images.
The distance ratio  is a ratio of angular distances, 
\begin{equation}
 D\equiv \frac{D_{OL}D_{OS} }{D_{LS}}\,,
 \end{equation}
where the distance from the observer to the lens is $D_{OL}$, 
the distance from the observer to the source is $D_{OS}$, and the distance from the lens to the source is $D_{LS}$.
We calculate distances based on a homogenous Friedmann-Lema{\^i}tre-Robertson-Walker cosmology,
with H$_0=h\times$~100~km/s/Mpc, where $h=0.673$, 
the mean mass density $\Omega_M=0.315$ and the normalized 
cosmological constant $\Omega_\Lambda=0.686$ \citep{2013arXiv1303.5076P}.

The magnification ratio and the time delay both change with the distance between the emitting regions in the source plane.
Differences in the projected orientation of the jet in the source plane also impact  the magnification ratio and time delay. 

As specific examples of the effects that source structure and orientation can have on H$_0$, 
we first consider five different jet locations and alignments in the source plane (Figure~\ref{fig:image_planeSIS}). 
We model the lens as a singular isothermal sphere and we compute differences in magnification ratios 
and time delays as a function of  projected distance between the emitting regions.
We then demonstrate  the impact of these variations  on the estimation of H$_0$. 

\section{Strong Gravitational Lensing}
\label{sec:SL}

\begin{figure}
\begin{center}
\includegraphics[width=6.2cm,angle=-90]{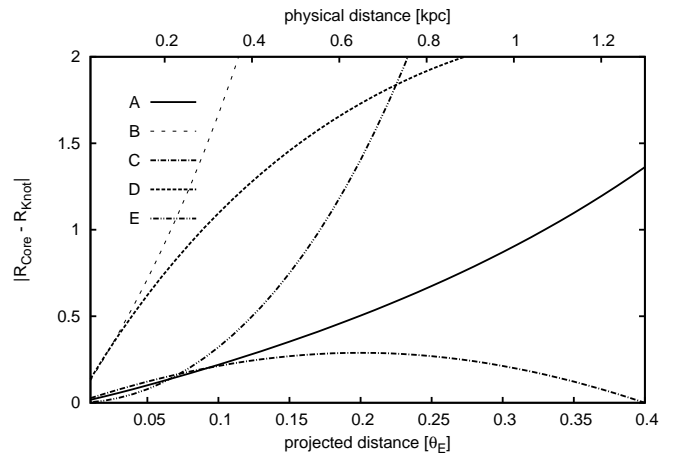}
\end{center}
\caption{\label{fig:mr_vs_pd_SIS} Difference in magnification ratio between the core and the knot as a function of projected distance between them. 
						     The figure shows the absolute value of the difference in the magnification ratio.
						     The projected distance between the core and the knots is presented in Einstein radius units (bottom),
						     and in physical units corresponding to the source plane (top).}
\end{figure}

\begin{figure}
\begin{center}
\includegraphics[width=6.2cm,angle=-90]{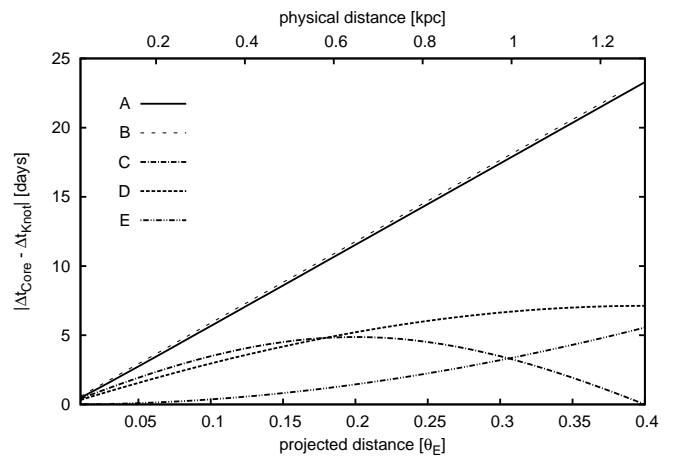}
\end{center}
\caption{\label{fig:dt_vs_pd_SIS} Difference in time delay between the core and the knot along the jet as a function of projected distance between them.}
\end{figure}

 \begin{figure}
\begin{center}
\includegraphics[width=6.2cm,angle=-90]{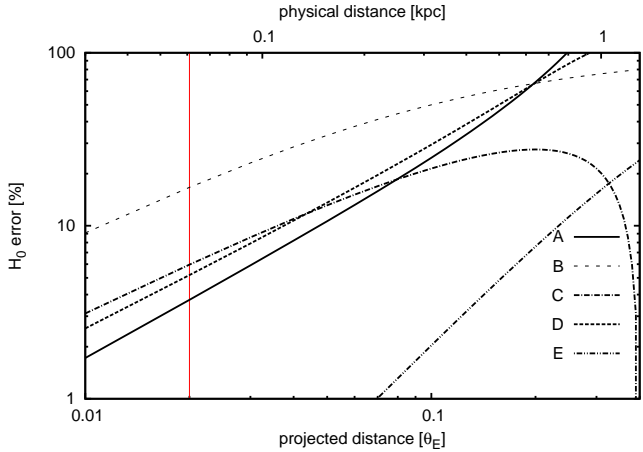}
\end{center}
\caption{\label{fig:h_vs_pd_SIS} Deviation of the derived Hubble parameter as a function of the projected distance 
						    between the core and the emitting region within the jet. 
						    The error is calculated as (H-H$_0$)/H$_0$,
						    where H$_0$ is a real value of the Hubble constant.
						    H is the value of the Hubble constant calculated using the  position of the core, 
						    and the time delay originating from the emitting region along the jet at a given projected distance from the core.
						    The perpendicular line indicates the projected distance between the core and HST-1 observed in M87.}
\end{figure}

Here we investigate the five examples of Figure~\ref{fig:image_planeSIS} in detail. 
We review the elements of the lensing model and we demonstrate the substantial effects of source structure on the
value of H$_0$ derived from the measured time delay.

Light rays from a source in the presence  of matter are deflected by an angle $\alpha$ 
before reaching an observer. 
In the thin-lens approximation, the rays are deflected by an angle of
\begin{equation}
\vec{\alpha}=\bigtriangledown \psi(\vec{\theta}) \,,
\label{eq:alpha}
\end{equation}
where  $\psi(\vec{\theta})$ is the effective gravitational potential of the lens at an image position  $\vec{\theta}$. 

Using simple geometry \citep[see][]{1996astro.ph..6001N,2013arXiv1307.4050B},  the lens equation is
 \begin{equation}
\vec{\beta}=\vec{\theta} - \vec{\alpha}(\vec{\theta}) \,,
\label{eq:beta}
\end{equation}
where $\vec{\beta}$ is the position of a source in the image plane.
The lens equation is generally not linear.
Thus, multiple images, $\vec{\theta_{\pm}}$ , 
corresponding to a single source position can be produced.

The time delay function  can be obtained from Fermat's principle:
\begin{equation}
\frac{c\, \Delta t(\vec{\theta})}{(z_L+1)} =
D\left [  \frac{1}{2}(\vec{\theta} -\vec{\beta})^2 -\psi(\vec{\theta})\right ]  \,,
 \label{dt3}
\end{equation}
where $\Delta t$ is the time delay, and $c$ is the speed of light.
The distance ratio $D$ is inversely proportional to  $H_0$, 
thus equation~(\ref{dt3}) provides a direct route to the Hubble constant parameter 
once the time delays are measured.

The time delay is proportional to the square of the angular offset between $\vec{\theta}$ and $\vec{\beta}$.
Therefore, different time delays result when  emission originates from different regions,
for example, a core or knots along a jet.

Light deflection in a gravitational field not only delays the rays, but  also magnifies the sources.
These magnifications are given by the determinant of the Jacobian matrix of the lens mapping $\theta \rightarrow \beta$.
The magnification factor is
\begin{equation}
A=\left| \mbox{det} \frac{\partial \beta}{\partial \theta} \right|^{-1} \,.
\end{equation}

The magnification $A$ is the ratio between  the flux of an image and the flux of the unlensed source.
If a source is mapped onto several images, the ratios of the respective magnification factors are equal 
to the flux ratios of the images \citep{1992grle.book.....S}.

We use the singular isothermal sphere (SIS) profile to characterize the lens.
This model is the simplest parameterization of 
the spatial distribution of matter in astronomical systems 
like galaxies and clusters of galaxies.
In general, the mass distribution of the lens composed of  stellar and dark matter 
is well represented by an isothermal model over many orders of magnitude in radius,
and deviate significantly only far outside the Einstein radius where it is not relevant for our results. 
The SIS model is consistent with the results 
of the Sloan Lens ACS Survey \citep{2007ApJ...667..176G,2010ApJ...724..511A}.

The lensing potential of the SIS is defined as
\begin{equation}
\psi(\theta)=\frac{D_{LS}}{D_{OS}} \frac{4\pi \sigma^2}{c^2}|\theta| \,,
\end{equation}
where $\sigma$ is a velocity dispersion.

Using this lensing potential,  equations~(\ref{eq:alpha}) and~(\ref{eq:beta}), 
the image position is
\begin{equation}
\label{eq:theta}
\theta_{\pm} = \beta\pm\theta_E \,,
\end{equation}
where $\theta_E$ is the Einstein radius.

The magnification ratio between images, $A_{\pm}$, is
\begin{equation}
R=\frac{A_+}{A_-} = \frac{\theta_+}{\theta_-} \,,
\end{equation}

and the time delay for $\beta>0$ is

\begin{equation}
\label{eq:tdSIS}
\frac{c\Delta t}{(1+z_L)} =  D \theta_E(\theta_E - (|\theta_+| - |\theta_-|)) \,.
\end{equation}

Figure~\ref{fig:mr_vs_pd_SIS} shows
the difference in the magnification ratio between the core and the emitting region along the jet
as a function of the projected distance between them.
Figure~\ref{fig:dt_vs_pd_SIS}  shows the difference in the time delays for the same configurations. 
The complex structure of the source produces significant differences  in both the time delays and magnification ratios.  
For example,  emitting regions separated by a projected distance of 5\% of the Einstein radius produce a difference in the magnification ratios in the range 0.1 to 0.7 
depending on the alignment in the source plane.
In our model 5\% of the Einstein radius corresponds to a projected distance of $\sim 160\,$pc. 
For the same projected distance the change in the time delay ranges from 1 to 3 days.

The rate of change of the lens parameters depends on the jet alignment in the source plane 
(see Figures~\ref{fig:mr_vs_pd_SIS} and~\ref{fig:dt_vs_pd_SIS}).
The largest rate of change in time delay and magnification ratios occurs
when the jet is oriented  along the radial direction in the source plane.

Figure~\ref{fig:h_vs_pd_SIS} shows the absolute value of the normalized difference between 
the true Hubble parameter H$_0$,
and  the value of the Hubble parameter, H,  calculated using the  position of the core 
and the time delay originating from the emitting region along the jet.
The departures from the true value  
arise from  misinterpreting the location of the variable emitting region. In other words, they arise from the assumption that the variable emitting region is always the same and that it is the core.  
Even for a small separation between the true variable emission region and the core of $\sim$60~parsecs,
the deviation between the derived Hubble parameter and the true value can be as large as 20\%. 

For powerful jets, 
the projected separation between the core and an emitting regions along the jet can 
be as large as hundreds of kpcs.
Note that for some jet alignments (e.g. A and D), 
the deviation of the H$_0$ measurement from the true value can become as large as 100\% 
for separations of 20\% of the Einstein radius. In the
toy model  $\sim$600~parsecs is $\sim$20\% of the Einstein radius.  

\section{Monte Carlo Simulations} 
\label{sec:MC}

The five examples we have explored thus far demonstrate the remarkable impact of 
source structure on the Hubble parameter derived from time delays of strongly-lensed sources. 
The next goal is to compare the basic model with existing observations.

In this section, we explore the existing measurements of H$_0$  determined from strongly-lensed sources. 
We take the measurements at face value and we compare the distribution
of these measured values with a Monte Carlo simulation that samples the full source plane for 
a set of sources with a variety of projected distances between the varying emission region and the core.

Figure~\ref{fig:h_data} shows a compilation of the measurements (Table~\ref{tab:H0}). 
Note the broad distribution of the individual measurements 
scattering in the range from 40 to 100. 
The distribution of  H$_0$ shows an underrepresented number of measurements 
around 67. This value  corresponds to  the H$_0$ reported by Planck \citep{2013arXiv1303.5076P}.
The distribution has  preferred values below 64, 
and broad scatter for values above 70. 

\subsection{Model} 
\label{sec:mc:sim}

\begin{figure}
\begin{center}
\includegraphics[width=9cm,angle=0]{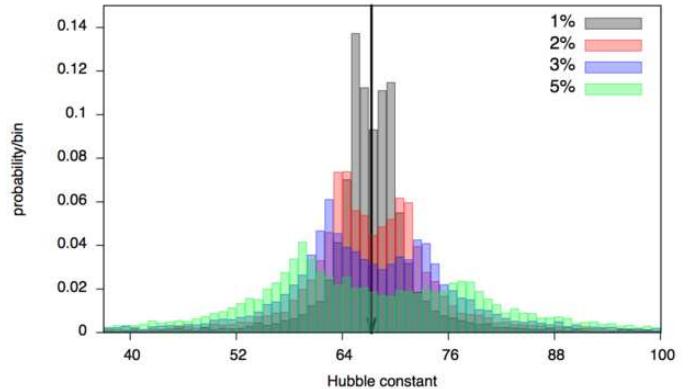}
\end{center}
\caption{\label{fig:h_MC} Distributions of the Hubble constant obtained with the Monte Carlo simulation 
                                         assuming different distances between the quasar and the emitting region along the jet.  
                                         Black line indicates the Planck satellite estimation of 67.3. }
\end{figure}

We ask whether a model based on simple assumptions about the structure of the sources
can reproduce the salient features of the observed distribution of lensing H$_0$ values.
In the simplest model, the location
of a quasar and the alignment of its jet are both random in the source plane.
We perform Monte Carlo simulations by randomly selecting the position of a quasar in the source plane. 
For each quasar position, we  also randomly select the jet alignment.
Then, we compute  H$_0$  for a source located at the position of the quasar and we compute 
the time delay that corresponds to a variable emission region positioned along the jet, rather than in the core. 
Figure~\ref{fig:h_MC} shows the distributions of H$_0$ for   
distances between the quasar and variable emitting region along the jet 
that correspond to 1\%, 2\%, 3\% and 5\% of the Einstein radius. 
We simulate $10^5$ sources for each separation.

We assume  an H$_0$ of 67.3 to calculate  time delays.
The recovered average value of  H$_0$ from the distributions  is $\sim 67.2 - 67.4$. 
The effect of the misinterpreted position of the source  cancels  out in the mean when 
large numbers of measurements from gravitationally-lensed systems are averaged. 
For any individual source, the deviation from the true value can be large.
 
\subsection{Bimodal Distribution} 
\label{sec:mc:sim}
 
The simulated distribution of H$_0$ has a distinctive  bimodal character.
This bimodality   is a consequence of the geometry, 
and the change of lens parameters across the source plane.
 The position of  the images and the time delay change with the distance from the  center of the lens mass distribution
 (see equations~(\ref{eq:theta}) and~(\ref{eq:tdSIS})).
The alignment of the jet  in the source plane is random.
The distance of the variable emitting region from the center of the lens mass distribution can be either smaller or larger 
than the distance between the quasar and the center of the lens mass distribution.
Thus, the values of H$_0$ are  either over or underestimated (Figures~\ref{fig:scheme1} and~\ref{fig:scheme2}).

 \begin{figure}
\begin{center}
\includegraphics[width=9cm,angle=0]{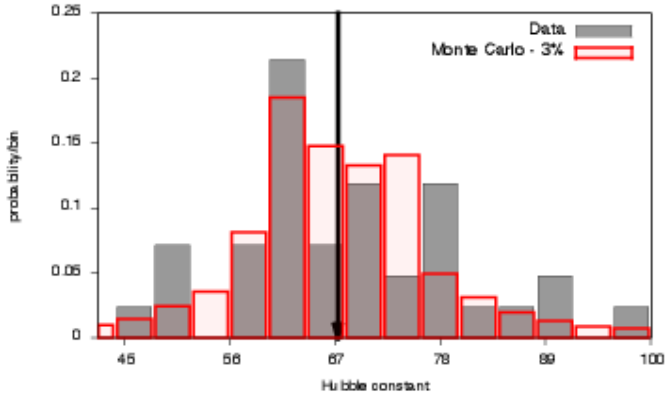}
\end{center}
\caption{\label{fig:h_data} Gray distribution represents Hubble constant measurements  summarized  in  table~\ref{tab:H0}.
				          Red distribution is  the results of Monte Carlo simulations of 4000 trials, 
				          assuming distance between the core and variable emitting region of $3\%$ of Einstein radius.
				          Black line indicates the Planck satellite estimation of 67.3. }
\end{figure}

The chance that the jet will be aligned in such a way that the emitting region along the jet and the core 
are at the same distance from the center of the lens  is small. 
Therefore, the simulated distribution of  H$_0$ shows that  the nominal value is underrepresented in the distribution. 
On other words, there is a dip in the center of the distribution of recovered values of H$_0$.

The Root-Mean-Square (RMS) of the H$_0$ distribution increases with  projected separation between the variable emission source and the core. 
For distances between the core and an emission region along the jet corresponding to 
1\%,2\%, 3\% and 5\% of the Einstein radius, 
the corresponding spread in the Hubble parameter is 5.65, 7.92, 9.56, and 12.29  [$\mbox{km\,s}^{-1}$Mpc$^{-1}$], respectively.
The scatter in H$_0$ reflects the offset between the position of the variable emitting region and the assumed position of the core. 
Thus, the RMS of measured H$_0$ together with true value of H$_0$ determine the most likely offset between 
the core and the variable emitting region.

\subsection{Position of the Source} 
\label{sec:mc:ps}

\begin{figure}
\begin{center}
\includegraphics[width=8.5cm,angle=0]{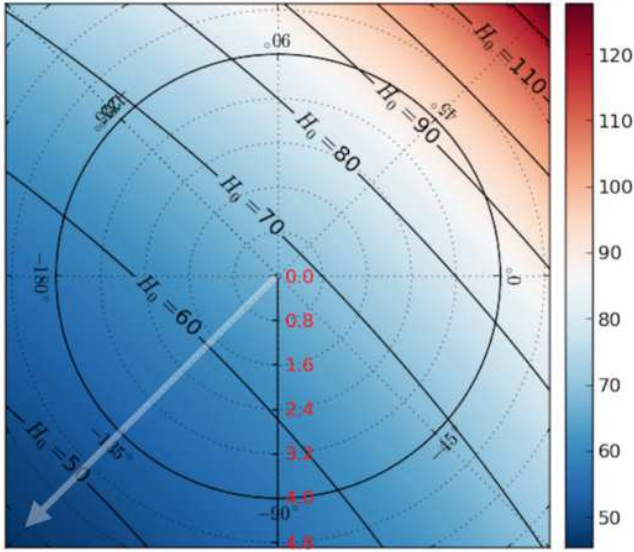}
\end{center}
\caption{\label{fig:scheme1} A schematic view showing how determination of the Hubble parameter depends 
					on the reconstruction of the location of the variable emitting  region.
					The light arrow indicates the direction towards the center of the lens mass.
					In this example the source is located at 0.35 Einstein  radius from the  center of the lens mass. 
					An auxiliary  grid is centered at the position of the core. 
					The angular coordinates show the alignment of the jet in the source plane relative to the position of the core.
					The radial coordinates indicate the distance between the core and the variable emitting region along to the jet,
					shown as a percentage of the Einstein radius.
					The contour map shows the value of the Hubble parameter obtained by assuming the position of 
					the core at 0.35 Einstein  radius from the  mass center  of the lens,
					and time delays  corresponding to variable emitting regions at the indicated distances from the core. }
\end{figure}

\begin{figure}
\begin{center}
\includegraphics[width=8.5cm,angle=0]{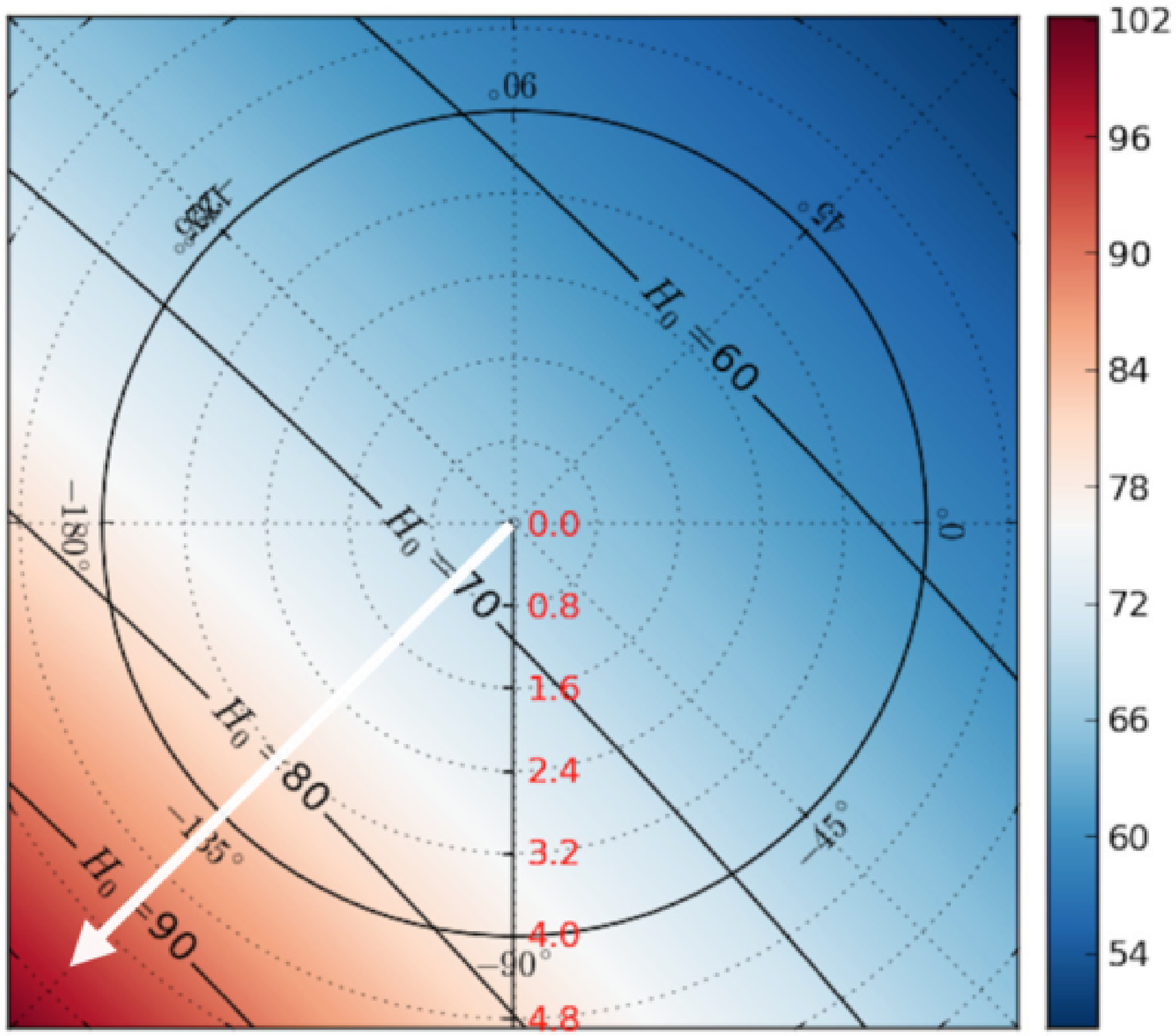}
\end{center}
\caption{\label{fig:scheme2} The same as Figure~\ref{fig:scheme1}, 
					   but with the core located at 0.7 Einstein radius from the  mass center of the lens. }
\end{figure}

The value of H$_0$ is clearly very sensitive to the position of the source. 
Figures~\ref{fig:scheme1} and~\ref{fig:scheme2} show examples 
where the core is located at 0.35 Einstein radius and 0.7 Einstein radius from the  center of the lens mass, respectively. 
The scale shows how the estimated Hubble parameter changes when the variable emission originates at a  particular distance from the core. 
Figures~\ref{fig:scheme1} and~\ref{fig:scheme2} explore  distances between the core and emitting region along the jet only up to $\sim5\,$\% of the Einstein radius.
Even for very small distances, like 2.5\% of Einstein radius,  
differences in the derived Hubble parameter can be large, spanning the range 55 - 80.

The reconstructed value of the Hubble parameter depends on the  alignment of the jet,
which can be reconstructed from radio images showing a number of lensed features.

The positions of the images of the core are resolved with optical and radio instruments.
To identify the source, at least one resolved image is necessary.
In the case of gamma-ray observations, the time delay can be measured with a precision of a few percent 
even when the images are poorly resolved or unresolved. The true source of the variable emissions generally unknown. 
Figures~\ref{fig:scheme1} and \ref{fig:scheme2} demonstrate that changes in the distance between source of variable emission and the core and/or a small error in the positions of a variable core can lead to wide variation in the derived Hubble parameter.

\section{Discussion} 
\label{sec:dis}

\subsection{Comparison with Lensing Measurements of H$_0$} 
\label{sec:dis:H0}

Time delays between the mirage images of gravitationally lensed variable sources 
are exploited  to estimate the Hubble parameter.
The sources which exhibit the most prominent variability over short time periods,
and across the entire electromagnetic spectrum, 
are objects with jets, including, for example, blazars \citep{1997ARA&A..35..445U,2012ApJ...760...51R}. 

Figure~\ref{fig:h_data} shows the distribution (gray) 
of 37 measurements of H$_0$ using gravitationally lensed systems available in the literature (Table~\ref{tab:H0}).
This distribution is  characterized by a mean value of 67.95 with an RMS scatter of 12.4 [$\mbox{km\,s}^{-1}$Mpc$^{-1}$].

The red distribution represents the Monte Carlo simulations composed of 4000 trials. 
For this comparison, we take an offset between the position of the source and the position of the variable emitting region 
equal to 3\% of the Einstein radius. 
The resulting simulated distribution is characterized by a mean value 67.3 with an RMS
scatter of 9.6  [$\mbox{km\,s}^{-1}$Mpc$^{-1}$]. Even for this very simple model, 
the uncertainty in the reconstruction of the position of the variable emitting region 
reproduces the scatter and the features in the H$_0$ measurement distribution obtained from observed gravitationally lensed systems. 

An uncertainty in  the position of the source in the source plane larger than 1\% of the Einstein radius
can lead to the same effect as the complex structure of the source, 
even when the variable emitting region and resolved images spatially coincide (see Figure~\ref{fig:h_MC}). 
Note that  a typical Einstein radius for these sources is $\sim 0.5\, $arcsecond,
and 5\%  corresponds to 0.025 arcseconds, only twice the HST resolution. 
Even with HST images, there is a natural limit to reconstruction that can produce a spread in the derived values of H$_0$.

Table~\ref{tab:H0} contains a compilation of the H$_0$ measurements available in the literature.
These measurements were obtained using different lens modeling, 
  treatments of the lens environments,
and different methods  for  time delay measurements,
with some of the time delays measurement being disputed \citep{2011A&A...536A..44E}.
In the future, we expect these effects will be minimized as projects like 
COSMOGRAIL \citep{2005A&A...436...25E,2006A&A...451..747E,2006A&A...451..759E,2006A&A...450..461S,2007A&A...464..845V,
2007A&A...465...51E,2008A&A...488..481V,2010A&A...522A..95C,2011A&A...536A..53C,2012A&A...538A..99S,2013A&A...553A.120T,
2013A&A...553A.121E,2013A&A...556A..22T,2013A&A...557A..44R}
provide uniform analysis and data for large sample.

All of these differences can contribute significantly to the scatter in H$_0$ in Table~\ref{tab:H0}.
For demonstration purposes of this paper, 
we take measurements of H$_0$ at face value. 
Of course, the other sources of error will change the uncertainty in
variability position required to fit the data, but in this work we are
concerned with showing what the effect of source variability
positional uncertainty is, rather than with quantifying its true
contribution to the scatter in H$_0$ measurements.

The complex structure of the source predicts a very characteristic bimodal distribution,
with a dip at the true value. 
Other systematics may manifest their presence in the distribution of the Hubble constant  in different ways.
They may shift or broaden the distribution.
Thus, in the future, when  large samples of H$_0$ measurements become available, 
the shape of the observed H$_0$ distribution 
will be a useful tool for identifying and quantifying these systematics. 

Currently  $\sim200$ gravitationally lensed systems have been identified. 
Among them, time delays have been estimated  for $\sim20$ systems. 
Future survey instruments like SKA, LSST or Euclid, 
will increase the number of known gravitationally lensed systems by about two orders of magnitudes.  
Predictions suggest that  LSST alone will provide $\sim4000$ time delays \citep{2009ApJ...706...45C}.
The distribution of Hubble parameters obtained from these large ensembles of time delays 
will allow tests of these model of the impact of the complex structure on the derived Hubble parameter. They will also provide a statistical determination of H$_0$ that can take the complex source structure into account.

\subsection{The Hubble Parameter Tuning Approach} 
\label{sec:dis:tuneH0}

The measured value of the Hubble parameter is very sensitive to the spatial offset between the 
position of the core and the position  of the variable-emitting region
where the time delay originates. 
Thus, the problem can be inverted. 
The value of H$_0$ measured by  other techniques
can be used to tune the spatial offset between the position of well-resolved images of lensed jets
and the position of the variable-emitting region where the time delay originates.
We call this approach the Hubble Parameter Tuning (HPT) method for spatially resolving the active region.

The four key components of the HPT method  are: 
the value of the H$_0$, 
a model for the lens, 
the resolved positions of  emitting regions,
and  time delays measured at high energies. 
We consider each of these issues next.

Currently, various  methods of  H$_0$ estimation measure the value of H$_0$ with an accuracy down to $\sim$2\%.
However, accounting for the tension between different methods,  H$_0$ is measured with 10\% accuracy.  
In the near future, these methods have the potential for revealing the value of H$_0$  to 1\% accuracy 
\citep[for a brief review see][]{2012arXiv1202.4459S}.

Modeling of the mass distribution of the lens may be limited by, for example, mass-sheet degeneracy \citep{2013A&A...559A..37S}.
Reconstruction of the mass distribution of the lens, in some cases, can be  the most constraining element in resolving the spatial origin 
of variable emission.
However, the images of   lensed jets are observed with high accuracy from radio, through optical, up to even x-rays with Chandra.
The images of emitting regions  are resolved at radio with an angular resolution better than 1~milliarcsecond.
The morphology of these images is very complex, showing  ring structures and multiple images\footnote{http://www.jb.man.ac.uk/research/gravlens/lensarch/lens.html}.
Thus, lensed jets  can help to better constrain the mass distribution of the lens.

Emission from  jets can increase by an order of magnitude in a short period of time.
The variability time scales (minutes to days) are much shorter than 
the typical time delays (weeks  to years).
As a result, time delays can be measured despite poorly resolved, or even unresolved, images.

The HPT approach provides an opportunity, 
to identify the origin of the variable emission. 
The answer to this question is crucial at energies where 
current detectors cannot resolve the sources. 

\subsection{Gamma-Ray and X-Ray Observations} 
\label{sec:dis:gamma}

Emission from extragalactic variable sources obviously cannot be resolved at gamma-ray energies; 
the angular resolution of current detectors is limited to $\sim0.1\,$degree.
In this energy range,  improvement by a factor of 1000 in angular resolution is required to 
determine whether 
the flaring gamma-ray activity originates from  a  region close to the core or from knots along the jet. 

Future gamma-ray detectors  will provide an improvement in  angular resolution by 
a factor of a few \citep{2014arXiv1406.4830B,2014arXiv1407.0710W}.
Further improvements  are physically limited by effects like  nuclear recoil. 
Thus, gamma-ray detectors  with a galaxy acting as a lens located along the line of sight to a blazar
offer  the only way to answer the question of the origin of gamma-ray flares.  

In the x-ray, Chandra has an angular resolution of $\sim 0.5$ arcseconds. 
The angular resolution in the x-ray allowed  the 
discovery that extragalactic jets have complex x-ray structure,
and that the x-ray emission can extend over hundreds of kpcs. 
The excellent angular resolution of the Chandra instrument enabled the discovery
that  x-ray flares can originate from distant knots \cite[see example of M87 in][]{2006ApJ...640..211H}. 

Even in the x-ray,  resolving the sources becomes challenging for more distant sources.
Thus, gravitationally lensed x-ray sources offer a way to resolve the emission from jets 
associated with sources located at high redshift. 
The study of lensed sources provides a route to  investigation of the cosmic evolution of jets.

X-ray detectors like NuSTAR or {\it Swift}/XRT have an angular resolution of a few arcseconds.
Monitoring of x-ray variable gravitationally lensed sources with these instruments 
can enhance the ability of these instruments to  resolve the emission on scales as small as 
0.01 arsecend, depending on the properties of the particular lens system. 
%

\section{Conclusions}
\label{sec:con}

We use  Monte Carlo simulations where we
randomly select the position of the core and the jet alignments in the source plane 
to predict the distribution of values of H$_0$ obtained from complex lensed variable sources.
The only free parameter in our model is the separation between the variable emission region and the core.
We investigate the  impact of the misinterpreted position of 
the variable emitting region on the distribution of  H$_0$.

The Monte Carlo simulations successfully reproduce the character of the  observed  distribution 
of H$_0$ derived from individual gravitationally lensed systems. 
Misalignment by  5\% of Einstein radius results in a bimodal  distribution 
of values of H$_0$ characterized by an RMS of $\sim12$ [$\mbox{km\,s}^{-1}$Mpc$^{-1}$],
similar to the observed RMS of 12.4 [$\mbox{km\,s}^{-1}$Mpc$^{-1}$].  
Estimation of the Hubble parameter from strongly lensed sources 
is very sensitive to the position of the region in the lens plane where  variability originates. 
Nonetheless the mean Hubble parameter for a large sample returns the true value.

Other methods determine H$_0$  to an accuracy of 2\%.
Thus, we propose taking the value of H$_0$ 
from these techniques as given and then  using the distribution of H$_0$ values for gravitationally lensed systems 
to locate  the variable emission regions. In other words, taking H$_0$ as a known quantity
enables the use of gravitational lensing to effectively enhance the resolution of distant
variable sources. We call this resolution enhancing approach
the Hubble parameter tuning method for spatially resolving the source of variable emission.   

This H$_0$ tuning approach can be especially useful at energies 
where the emission from extragalactic sources cannot be spatially resolved 
due to a poor angular resolution of the detectors. 
In particular, it can be used at gamma rays where detectors  with a galaxy acting as 
a lens located along the line of sight to a blazar
offer  the only way to answer the question of the origin of gamma-ray flares.  
In the x-ray the resolution of current detectors is adequate to resolve nearby extragalactic sources, 
but application of techniques based on strong lensing enables extension of our knowledge of the structure of these sources to much greater redshift. 
The method can also be  used to enhance  the spatial resolution of variable emission even in the optical and NIR,
when high resolution images are not available during an outburst, or when the source is located at very high redshift.     
The extension to large redshift opens the possibility of following the evolution of the source structure with cosmic time,
especially when large samples from projects like Euclid and SKA are available.

\acknowledgments
We thank the referee for  valuable comments that prompted improvements in the manuscript. 
We thank Markus B\"ottcher, Dominique Sluse, Tommaso Treu, 
and Jabran Zahid for the valuable comments on the manuscript.
A.B. is supported by the Department of Energy Office of Science, 
NASA \& the Smithsonian Astrophysical Observatory 
with financial support from the NCN grant DEC-2011/01/M/ST9/01891.
MJG is supported by the Smithsonian Institution.

\bibliography{H0_blazars_v1}

\begin{thebibliography}{}
\expandafter\ifx\csname natexlab\endcsname\relax\def\natexlab#1{#1}\fi

\bibitem[{{Abramowski} {et~al.}(2012){Abramowski}, {Acero}, {Aharonian},
  {Akhperjanian}, {Anton}, {Balzer}, {Barnacka}, {Barres de Almeida},
  {Becherini}, {Becker}, \& et~al.}]{2012ApJ...746..151A}
{Abramowski}, A., {Acero}, F., {Aharonian}, F., {et~al.} 2012, \apj, 746, 151

\bibitem[{{Aharonian} {et~al.}(2006){Aharonian}, {Akhperjanian}, {Bazer-Bachi},
  {Beilicke}, {Benbow}, {Berge}, {Bernl{\"o}hr}, {Boisson}, {Bolz}, {Borrel},
  {Braun}, {Brown}, {B{\"u}hler}, {B{\"u}sching}, {Carrigan}, {Chadwick},
  {Chounet}, {Coignet}, {Cornils}, {Costamante}, {Degrange}, {Dickinson},
  {Djannati-Ata{\"i}}, {Drury}, {Dubus}, {Egberts}, {Emmanoulopoulos},
  {Espigat}, {Feinstein}, {Ferrero}, {Fiasson}, {Fontaine}, {Funk}, {Funk},
  {F{\"u}{\ss}ling}, {Gallant}, {Giebels}, {Glicenstein}, {Goret},
  {Hadjichristidis}, {Hauser}, {Hauser}, {Heinzelmann}, {Henri}, {Hermann},
  {Hinton}, {Hoffmann}, {Hofmann}, {Holleran}, {Hoppe}, {Horns},
  {Jacholkowska}, {de Jager}, {Kendziorra}, {Kerschhaggl}, {Kh{\'e}lifi},
  {Komin}, {Konopelko}, {Kosack}, {Lamanna}, {Latham}, {Le Gallou},
  {Lemi{\`e}re}, {Lemoine-Goumard}, {Lenain}, {Lohse}, {Martin},
  {Martineau-Huynh}, {Marcowith}, {Masterson}, {Maurin}, {McComb}, {Moulin},
  {de Naurois}, {Nedbal}, {Nolan}, {Noutsos}, {Orford}, {Osborne}, {Ouchrif},
  {Panter}, {Pelletier}, {Pita}, {P{\"u}hlhofer}, {Punch}, {Ranchon},
  {Raubenheimer}, {Raue}, {Rayner}, {Reimer}, {Ripken}, {Rob}, {Rolland},
  {Rosier-Lees}, {Rowell}, {Sahakian}, {Santangelo}, {Saug{\'e}}, {Schlenker},
  {Schlickeiser}, {Schr{\"o}der}, {Schwanke}, {Schwarzburg}, {Schwemmer},
  {Shalchi}, {Sol}, {Spangler}, {Spanier}, {Steenkamp}, {Stegmann}, {Superina},
  {Tam}, {Tavernet}, {Terrier}, {Tluczykont}, {van Eldik}, {Vasileiadis},
  {Venter}, {Vialle}, {Vincent}, {V{\"o}lk}, {Wagner}, \&
  {Ward}}]{2006Sci...314.1424A}
{Aharonian}, F., {Akhperjanian}, A.~G., {Bazer-Bachi}, A.~R., {et~al.} 2006,
  Science, 314, 1424

\bibitem[{{Auger} {et~al.}(2010){Auger}, {Treu}, {Bolton}, {Gavazzi},
  {Koopmans}, {Marshall}, {Moustakas}, \& {Burles}}]{2010ApJ...724..511A}
{Auger}, M.~W., {Treu}, T., {Bolton}, A.~S., {et~al.} 2010, \apj, 724, 511

\bibitem[{{Barnacka}(2013)}]{2013arXiv1307.4050B}
{Barnacka}, A. 2013, ArXiv e-prints, arXiv:1307.4050

\bibitem[{{Barnacka} {et~al.}(2014{\natexlab{a}}){Barnacka}, {B{\"o}ttcher}, \&
  {Sushch}}]{2014arXiv1404.4422B}
{Barnacka}, A., {B{\"o}ttcher}, M., \& {Sushch}, I. 2014{\natexlab{a}}, \apj,
  790, 147

\bibitem[{{Barnacka} {et~al.}(2014{\natexlab{b}}){Barnacka}, {Geller},
  {Dell'antonio}, \& {Benbow}}]{2014arXiv1403.5316B}
{Barnacka}, A., {Geller}, M.~J., {Dell'antonio}, I.~P., \& {Benbow}, W.
  2014{\natexlab{b}}, \apj, 788, 139

\bibitem[{{Barnacka} {et~al.}(2011){Barnacka}, {Glicenstein}, \&
  {Moudden}}]{2011A&A...528L...3B}
{Barnacka}, A., {Glicenstein}, J.-F., \& {Moudden}, Y. 2011, \aap, 528, L3

\bibitem[{{Bernard} {et~al.}(2014){Bernard}, {Bruel}, {Frotin}, {Geerebaert},
  {Giebels}, {Gros}, {Horan}, {Louzir}, {Poilleux}, {Semeniouk}, {Wang},
  {Anvar}, {Atti{\'e}}, {Colas}, {Delbart}, {Sizun}, \&
  {G{\"o}tz}}]{2014arXiv1406.4830B}
{Bernard}, D., {Bruel}, P., {Frotin}, M., {et~al.} 2014, ArXiv e-prints,
  arXiv:1406.4830

\bibitem[{{Bernstein} \& {Fischer}(1999)}]{1999AJ....118...14B}
{Bernstein}, G., \& {Fischer}, P. 1999, \aj, 118, 14

\bibitem[{{Biggs} {et~al.}(1999){Biggs}, {Browne}, {Helbig}, {Koopmans},
  {Wilkinson}, \& {Perley}}]{1999MNRAS.304..349B}
{Biggs}, A.~D., {Browne}, I.~W.~A., {Helbig}, P., {et~al.} 1999, \mnras, 304,
  349

\bibitem[{{Biretta} {et~al.}(1999){Biretta}, {Sparks}, \&
  {Macchetto}}]{1999ApJ...520..621B}
{Biretta}, J.~A., {Sparks}, W.~B., \& {Macchetto}, F. 1999, \apj, 520, 621

\bibitem[{{Biretta} {et~al.}(1991){Biretta}, {Stern}, \&
  {Harris}}]{1991AJ....101.1632B}
{Biretta}, J.~A., {Stern}, C.~P., \& {Harris}, D.~E. 1991, \aj, 101, 1632

\bibitem[{{Bolton} {et~al.}(2008){Bolton}, {Treu}, {Koopmans}, {Gavazzi},
  {Moustakas}, {Burles}, {Schlegel}, \& {Wayth}}]{2008ApJ...684..248B}
{Bolton}, A.~S., {Treu}, T., {Koopmans}, L.~V.~E., {et~al.} 2008, \apj, 684,
  248

\bibitem[{{Burud} {et~al.}(2002){Burud}, {Courbin}, {Magain}, {Lidman},
  {Hutsem{\'e}kers}, {Kneib}, {Hjorth}, {Brewer}, {Pompei}, {Germany},
  {Pritchard}, {Jaunsen}, {Letawe}, \& {Meylan}}]{Burud2002}
{Burud}, I., {Courbin}, F., {Magain}, P., {et~al.} 2002, \aap, 383, 71

\bibitem[{{Chantry} {et~al.}(2010){Chantry}, {Sluse}, \&
  {Magain}}]{2010A&A...522A..95C}
{Chantry}, V., {Sluse}, D., \& {Magain}, P. 2010, \aap, 522, A95

\bibitem[{{Cheung} {et~al.}(2014){Cheung}, {Larsson}, {Scargle}, {Amin},
  {Blandford}, {Bulmash}, {Chiang}, {Ciprini}, {Corbet}, {Falco}, {Marshall},
  {Wood}, {Ajello}, {Bastieri}, {Chekhtman}, {D'Ammando}, {Giroletti}, {Grove},
  {Lott}, {Ojha}, {Orienti}, {Perkins}, {Razzano}, {Smith}, {Thompson}, \&
  {Wood}}]{2014ApJ...782L..14C}
{Cheung}, C.~C., {Larsson}, S., {Scargle}, J.~D., {et~al.} 2014, \apjl, 782,
  L14

\bibitem[{{Coe} \& {Moustakas}(2009)}]{2009ApJ...706...45C}
{Coe}, D., \& {Moustakas}, L.~A. 2009, \apj, 706, 45

\bibitem[{{Courbin}(2003)}]{2003astro.ph..4497C}
{Courbin}, F. 2003, ArXiv Astrophysics e-prints, astro-ph/0304497

\bibitem[{{Courbin} {et~al.}(2011{\natexlab{a}}){Courbin}, {Chantry}, {Revaz},
  {Sluse}, {Faure}, {Tewes}, {Eulaers}, {Koleva}, {Asfandiyarov}, {Dye},
  {Magain}, {van Winckel}, {Coles}, {Saha}, {Ibrahimov}, \&
  {Meylan}}]{Courbin2011}
{Courbin}, F., {Chantry}, V., {Revaz}, Y., {et~al.} 2011{\natexlab{a}}, \aap,
  536, A53

\bibitem[{{Courbin} {et~al.}(2011{\natexlab{b}}){Courbin}, {Chantry}, {Revaz},
  {Sluse}, {Faure}, {Tewes}, {Eulaers}, {Koleva}, {Asfandiyarov}, {Dye},
  {Magain}, {van Winckel}, {Coles}, {Saha}, {Ibrahimov}, \&
  {Meylan}}]{2011A&A...536A..53C}
---. 2011{\natexlab{b}}, \aap, 536, A53

\bibitem[{{Eigenbrod} {et~al.}(2006{\natexlab{a}}){Eigenbrod}, {Courbin},
  {Dye}, {Meylan}, {Sluse}, {Vuissoz}, \& {Magain}}]{2006A&A...451..747E}
{Eigenbrod}, A., {Courbin}, F., {Dye}, S., {et~al.} 2006{\natexlab{a}}, \aap,
  451, 747

\bibitem[{{Eigenbrod} {et~al.}(2007){Eigenbrod}, {Courbin}, \&
  {Meylan}}]{2007A&A...465...51E}
{Eigenbrod}, A., {Courbin}, F., \& {Meylan}, G. 2007, \aap, 465, 51

\bibitem[{{Eigenbrod} {et~al.}(2006{\natexlab{b}}){Eigenbrod}, {Courbin},
  {Meylan}, {Vuissoz}, \& {Magain}}]{2006A&A...451..759E}
{Eigenbrod}, A., {Courbin}, F., {Meylan}, G., {Vuissoz}, C., \& {Magain}, P.
  2006{\natexlab{b}}, \aap, 451, 759

\bibitem[{{Eigenbrod} {et~al.}(2005){Eigenbrod}, {Courbin}, {Vuissoz},
  {Meylan}, {Saha}, \& {Dye}}]{2005A&A...436...25E}
{Eigenbrod}, A., {Courbin}, F., {Vuissoz}, C., {et~al.} 2005, \aap, 436, 25

\bibitem[{{Eulaers} \& {Magain}(2011)}]{2011A&A...536A..44E}
{Eulaers}, E., \& {Magain}, P. 2011, \aap, 536, A44

\bibitem[{{Eulaers} {et~al.}(2013){Eulaers}, {Tewes}, {Magain}, {Courbin},
  {Asfandiyarov}, {Ehgamberdiev}, {Rathna Kumar}, {Stalin}, {Prabhu}, {Meylan},
  \& {Van Winckel}}]{2013A&A...553A.121E}
{Eulaers}, E., {Tewes}, M., {Magain}, P., {et~al.} 2013, \aap, 553, A121

\bibitem[{{Fadely} {et~al.}(2010){Fadely}, {Keeton}, {Nakajima}, \&
  {Bernstein}}]{2010ApJ...711..246F}
{Fadely}, R., {Keeton}, C.~R., {Nakajima}, R., \& {Bernstein}, G.~M. 2010,
  \apj, 711, 246

\bibitem[{{Fassnacht} {et~al.}(1999){Fassnacht}, {Pearson}, {Readhead},
  {Browne}, {Koopmans}, {Myers}, \& {Wilkinson}}]{1999ApJ...527..498F}
{Fassnacht}, C.~D., {Pearson}, T.~J., {Readhead}, A.~C.~S., {et~al.} 1999,
  \apj, 527, 498

\bibitem[{{Fassnacht} {et~al.}(2002){Fassnacht}, {Xanthopoulos}, {Koopmans}, \&
  {Rusin}}]{2002ApJ...581..823F}
{Fassnacht}, C.~D., {Xanthopoulos}, E., {Koopmans}, L.~V.~E., \& {Rusin}, D.
  2002, \apj, 581, 823

\bibitem[{{Freedman} \& {Madore}(2010)}]{2010ARA&A..48..673F}
{Freedman}, W.~L., \& {Madore}, B.~F. 2010, \araa, 48, 673

\bibitem[{{Freedman} {et~al.}(2012){Freedman}, {Madore}, {Scowcroft}, {Burns},
  {Monson}, {Persson}, {Seibert}, \& {Rigby}}]{2012ApJ...758...24F}
{Freedman}, W.~L., {Madore}, B.~F., {Scowcroft}, V., {et~al.} 2012, \apj, 758,
  24

\bibitem[{{Freedman} {et~al.}(2001){Freedman}, {Madore}, {Gibson}, {Ferrarese},
  {Kelson}, {Sakai}, {Mould}, {Kennicutt}, {Ford}, {Graham}, {Huchra},
  {Hughes}, {Illingworth}, {Macri}, \& {Stetson}}]{2001ApJ...553...47F}
{Freedman}, W.~L., {Madore}, B.~F., {Gibson}, B.~K., {et~al.} 2001, \apj, 553,
  47

\bibitem[{{Gavazzi} {et~al.}(2007){Gavazzi}, {Treu}, {Rhodes}, {Koopmans},
  {Bolton}, {Burles}, {Massey}, \& {Moustakas}}]{2007ApJ...667..176G}
{Gavazzi}, R., {Treu}, T., {Rhodes}, J.~D., {et~al.} 2007, \apj, 667, 176

\bibitem[{{Gil-Merino} {et~al.}(2002){Gil-Merino}, {Wisotzki}, \&
  {Wambsganss}}]{Merino2002}
{Gil-Merino}, R., {Wisotzki}, L., \& {Wambsganss}, J. 2002, \aap, 381, 428

\bibitem[{{Harris} {et~al.}(2006){Harris}, {Cheung}, {Biretta}, {Sparks},
  {Junor}, {Perlman}, \& {Wilson}}]{2006ApJ...640..211H}
{Harris}, D.~E., {Cheung}, C.~C., {Biretta}, J.~A., {et~al.} 2006, \apj, 640,
  211

\bibitem[{{Harris} \& {Krawczynski}(2006)}]{2006ARA&A..44..463H}
{Harris}, D.~E., \& {Krawczynski}, H. 2006, \araa, 44, 463

\bibitem[{{Hjorth} {et~al.}(2002){Hjorth}, {Burud}, {Jaunsen}, {Schechter},
  {Kneib}, {Andersen}, {Korhonen}, {Clasen}, {Kaas}, {{\O}stensen}, {Pelt}, \&
  {Pijpers}}]{2002ApJ...572L..11H}
{Hjorth}, J., {Burud}, I., {Jaunsen}, A.~O., {et~al.} 2002, \apjl, 572, L11

\bibitem[{{Jakobsson} {et~al.}(2005){Jakobsson}, {Hjorth}, {Burud}, {Letawe},
  {Lidman}, \& {Courbin}}]{Jakobsson2005}
{Jakobsson}, P., {Hjorth}, J., {Burud}, I., {et~al.} 2005, \aap, 431, 103

\bibitem[{{Keeton} {et~al.}(2000){Keeton}, {Falco}, {Impey}, {Kochanek},
  {Leh{\'a}r}, {McLeod}, {Rix}, {Mu{\~n}oz}, \& {Peng}}]{2000ApJ...542...74K}
{Keeton}, C.~R., {Falco}, E.~E., {Impey}, C.~D., {et~al.} 2000, \apj, 542, 74

\bibitem[{{Kelly} {et~al.}(2009){Kelly}, {Bechtold}, \&
  {Siemiginowska}}]{2009ApJ...698..895K}
{Kelly}, B.~C., {Bechtold}, J., \& {Siemiginowska}, A. 2009, \apj, 698, 895

\bibitem[{{Kochanek}(2002)}]{2002ApJ...578...25K}
{Kochanek}, C.~S. 2002, \apj, 578, 25

\bibitem[{{Kochanek} \& {Schechter}(2004)}]{2004mmu..symp..117K}
{Kochanek}, C.~S., \& {Schechter}, P.~L. 2004, Measuring and Modeling the
  Universe, 117

\bibitem[{{Koopmans} {et~al.}(2000){Koopmans}, {de Bruyn}, {Xanthopoulos}, \&
  {Fassnacht}}]{Koopmans2000}
{Koopmans}, L.~V.~E., {de Bruyn}, A.~G., {Xanthopoulos}, E., \& {Fassnacht},
  C.~D. 2000, \aap, 356, 391

\bibitem[{{Koopmans} {et~al.}(2003){Koopmans}, {Treu}, {Fassnacht},
  {Blandford}, \& {Surpi}}]{2003ApJ...599...70K}
{Koopmans}, L.~V.~E., {Treu}, T., {Fassnacht}, C.~D., {Blandford}, R.~D., \&
  {Surpi}, G. 2003, \apj, 599, 70

\bibitem[{{Koz{\l}owski} {et~al.}(2010){Koz{\l}owski}, {Kochanek}, {Udalski},
  {Wyrzykowski}, {Soszy{\'n}ski}, {Szyma{\'n}ski}, {Kubiak}, {Pietrzy{\'n}ski},
  {Szewczyk}, {Ulaczyk}, {Poleski}, \& {OGLE
  Collaboration}}]{2010ApJ...708..927K}
{Koz{\l}owski}, S., {Kochanek}, C.~S., {Udalski}, A., {et~al.} 2010, \apj, 708,
  927

\bibitem[{{L{\"a}hteenm{\"a}ki} \& {Valtaoja}(1999)}]{1999ApJ...521..493L}
{L{\"a}hteenm{\"a}ki}, A., \& {Valtaoja}, E. 1999, \apj, 521, 493

\bibitem[{{Leh{\'a}r} {et~al.}(2000){Leh{\'a}r}, {Falco}, {Kochanek}, {McLeod},
  {Mu{\~n}oz}, {Impey}, {Rix}, {Keeton}, \& {Peng}}]{2000ApJ...536..584L}
{Leh{\'a}r}, J., {Falco}, E.~E., {Kochanek}, C.~S., {et~al.} 2000, \apj, 536,
  584

\bibitem[{{Lidman} {et~al.}(1999){Lidman}, {Courbin}, {Meylan}, {Broadhurst},
  {Frye}, \& {Welch}}]{1999ApJ...514L..57L}
{Lidman}, C., {Courbin}, F., {Meylan}, G., {et~al.} 1999, \apjl, 514, L57

\bibitem[{{Lobanov}(1998)}]{1998A&A...330...79L}
{Lobanov}, A.~P. 1998, \aap, 330, 79

\bibitem[{{MacLeod} {et~al.}(2010){MacLeod}, {Ivezi{\'c}}, {Kochanek},
  {Koz{\l}owski}, {Kelly}, {Bullock}, {Kimball}, {Sesar}, {Westman}, {Brooks},
  {Gibson}, {Becker}, \& {de Vries}}]{2010ApJ...721.1014M}
{MacLeod}, C.~L., {Ivezi{\'c}}, {\v Z}., {Kochanek}, C.~S., {et~al.} 2010,
  \apj, 721, 1014

\bibitem[{{Marscher}(2008)}]{2008ASPC..386..437M}
{Marscher}, A.~P. 2008, in Astronomical Society of the Pacific Conference
  Series, Vol. 386, Extragalactic Jets: Theory and Observation from Radio to
  Gamma Ray, ed. T.~A. {Rector} \& D.~S. {De Young}, 437

\bibitem[{{Marscher} {et~al.}(2008){Marscher}, {Jorstad}, {D'Arcangelo},
  {Smith}, {Williams}, {Larionov}, {Oh}, {Olmstead}, {Aller}, {Aller},
  {McHardy}, {L{\"a}hteenm{\"a}ki}, {Tornikoski}, {Valtaoja}, {Hagen-Thorn},
  {Kopatskaya}, {Gear}, {Tosti}, {Kurtanidze}, {Nikolashvili}, {Sigua},
  {Miller}, \& {Ryle}}]{2008Natur.452..966M}
{Marscher}, A.~P., {Jorstad}, S.~G., {D'Arcangelo}, F.~D., {et~al.} 2008, \nat,
  452, 966

\bibitem[{{Massaro} {et~al.}(2011){Massaro}, {Harris}, \&
  {Cheung}}]{2011ApJS..197...24M}
{Massaro}, F., {Harris}, D.~E., \& {Cheung}, C.~C. 2011, \apjs, 197, 24

\bibitem[{{Matthews} \& {Sandage}(1963)}]{1963ApJ...138...30M}
{Matthews}, T.~A., \& {Sandage}, A.~R. 1963, \apj, 138, 30

\bibitem[{{Narayan} \& {Bartelmann}(1996)}]{1996astro.ph..6001N}
{Narayan}, R., \& {Bartelmann}, M. 1996, ArXiv Astrophysics e-prints,
  astro-ph/9606001

\bibitem[{{Oguri}(2007)}]{2007ApJ...660....1O}
{Oguri}, M. 2007, \apj, 660, 1

\bibitem[{{Paraficz} \& {Hjorth}(2010)}]{2010ApJ...712.1378P}
{Paraficz}, D., \& {Hjorth}, J. 2010, \apj, 712, 1378

\bibitem[{{Perlman} {et~al.}(2011){Perlman}, {Adams}, {Cara}, {Bourque},
  {Harris}, {Madrid}, {Simons}, {Clausen-Brown}, {Cheung}, {Stawarz},
  {Georganopoulos}, {Sparks}, \& {Biretta}}]{2011ApJ...743..119P}
{Perlman}, E.~S., {Adams}, S.~C., {Cara}, M., {et~al.} 2011, \apj, 743, 119

\bibitem[{{Planck Collaboration} {et~al.}(2013){Planck Collaboration}, {Ade},
  {Aghanim}, {Armitage-Caplan}, {Arnaud}, {Ashdown}, {Atrio-Barandela},
  {Aumont}, {Baccigalupi}, {Banday}, \& et~al.}]{2013arXiv1303.5076P}
{Planck Collaboration}, {Ade}, P.~A.~R., {Aghanim}, N., {et~al.} 2013, ArXiv
  e-prints, arXiv:1303.5076

\bibitem[{{Pushkarev} {et~al.}(2010){Pushkarev}, {Kovalev}, \&
  {Lister}}]{2010ApJ...722L...7P}
{Pushkarev}, A.~B., {Kovalev}, Y.~Y., \& {Lister}, M.~L. 2010, \apjl, 722, L7

\bibitem[{{Rathna Kumar} {et~al.}(2014){Rathna Kumar}, {Stalin}, \&
  {Prabhu}}]{2014arXiv1404.2920R}
{Rathna Kumar}, S., {Stalin}, C.~S., \& {Prabhu}, T.~P. 2014, ArXiv e-prints,
  arXiv:1404.2920

\bibitem[{{Rathna Kumar} {et~al.}(2013){Rathna Kumar}, {Tewes}, {Stalin},
  {Courbin}, {Asfandiyarov}, {Meylan}, {Eulaers}, {Prabhu}, {Magain}, {Van
  Winckel}, \& {Ehgamberdiev}}]{2013A&A...557A..44R}
{Rathna Kumar}, S., {Tewes}, M., {Stalin}, C.~S., {et~al.} 2013, \aap, 557, A44

\bibitem[{{Refsdal}(1964)}]{1964MNRAS.128..307R}
{Refsdal}, S. 1964, \mnras, 128, 307

\bibitem[{{Rhee}(1991)}]{1991Natur.350..211R}
{Rhee}, G. 1991, \nat, 350, 211

\bibitem[{{Riess} {et~al.}(2011{\natexlab{a}}){Riess}, {Macri}, {Casertano},
  {Lampeitl}, {Ferguson}, {Filippenko}, {Jha}, {Li}, \&
  {Chornock}}]{2011ApJ...730..119R}
{Riess}, A.~G., {Macri}, L., {Casertano}, S., {et~al.} 2011{\natexlab{a}},
  \apj, 730, 119

\bibitem[{{Riess} {et~al.}(2011{\natexlab{b}}){Riess}, {Macri}, {Casertano},
  {Lampeit}, {Ferguson}, {Filippenko}, {Jha}, {Li}, {Chornock}, \&
  {Silverman}}]{2011ApJ...732..129R}
---. 2011{\natexlab{b}}, \apj, 732, 129

\bibitem[{{Ruan} {et~al.}(2012){Ruan}, {Anderson}, {MacLeod}, {Becker},
  {Burnett}, {Davenport}, {Ivezi{\'c}}, {Kochanek}, {Plotkin}, {Sesar}, \&
  {Stuart}}]{2012ApJ...760...51R}
{Ruan}, J.~J., {Anderson}, S.~F., {MacLeod}, C.~L., {et~al.} 2012, \apj, 760,
  51

\bibitem[{{Saha} {et~al.}(2006){Saha}, {Courbin}, {Sluse}, {Dye}, \&
  {Meylan}}]{2006A&A...450..461S}
{Saha}, P., {Courbin}, F., {Sluse}, D., {Dye}, S., \& {Meylan}, G. 2006, \aap,
  450, 461

\bibitem[{{Savolainen} {et~al.}(2010){Savolainen}, {Homan}, {Hovatta},
  {Kadler}, {Kovalev}, {Lister}, {Ros}, \& {Zensus}}]{2010A&A...512A..24S}
{Savolainen}, T., {Homan}, D.~C., {Hovatta}, T., {et~al.} 2010, \aap, 512, A24

\bibitem[{{Schechter}(2005)}]{2005IAUS..225..281S}
{Schechter}, P.~L. 2005, in IAU Symposium, Vol. 225, Gravitational Lensing
  Impact on Cosmology, ed. Y.~{Mellier} \& G.~{Meylan}, 281--296

\bibitem[{{Schechter} {et~al.}(1997){Schechter}, {Bailyn}, {Barr}, {Barvainis},
  {Becker}, {Bernstein}, {Blakeslee}, {Bus}, {Dressler}, {Falco}, {Fesen},
  {Fischer}, {Gebhardt}, {Harmer}, {Hewitt}, {Hjorth}, {Hurt}, {Jaunsen},
  {Mateo}, {Mehlert}, {Richstone}, {Sparke}, {Thorstensen}, {Tonry}, {Wegner},
  {Willmarth}, \& {Worthey}}]{1997ApJ...475L..85S}
{Schechter}, P.~L., {Bailyn}, C.~D., {Barr}, R., {et~al.} 1997, \apjl, 475, L85

\bibitem[{{Schneider} {et~al.}(1992){Schneider}, {Ehlers}, \&
  {Falco}}]{1992grle.book.....S}
{Schneider}, P., {Ehlers}, J., \& {Falco}, E.~E. 1992, {Gravitational Lenses}
  (Springer-Verlag Berlin Heidelberg New York)

\bibitem[{{Schneider} \& {Sluse}(2013)}]{2013A&A...559A..37S}
{Schneider}, P., \& {Sluse}, D. 2013, \aap, 559, A37

\bibitem[{{Schneider} \& {Sluse}(2014)}]{2014A&A...564A.103S}
---. 2014, \aap, 564, A103

\bibitem[{{Sereno} \& {Paraficz}(2014)}]{2014MNRAS.437..600S}
{Sereno}, M., \& {Paraficz}, D. 2014, \mnras, 437, 600

\bibitem[{{Sesar} {et~al.}(2007){Sesar}, {Ivezi{\'c}}, {Lupton}, {Juri{\'c}},
  {Gunn}, {Knapp}, {DeLee}, {Smith}, {Miknaitis}, {Lin}, {Tucker}, {Doi},
  {Tanaka}, {Fukugita}, {Holtzman}, {Kent}, {Yanny}, {Schlegel}, {Finkbeiner},
  {Padmanabhan}, {Rockosi}, {Bond}, {Lee}, {Stoughton}, {Jester}, {Harris},
  {Harding}, {Brinkmann}, {Schneider}, {York}, {Richmond}, \& {Vanden
  Berk}}]{2007AJ....134.2236S}
{Sesar}, B., {Ivezi{\'c}}, {\v Z}., {Lupton}, R.~H., {et~al.} 2007, \aj, 134,
  2236

\bibitem[{{Siemiginowska} {et~al.}(2002){Siemiginowska}, {Bechtold},
  {Aldcroft}, {Elvis}, {Harris}, \& {Dobrzycki}}]{2002ApJ...570..543S}
{Siemiginowska}, A., {Bechtold}, J., {Aldcroft}, T.~L., {et~al.} 2002, \apj,
  570, 543

\bibitem[{{Sluse} {et~al.}(2012){Sluse}, {Chantry}, {Magain}, {Courbin}, \&
  {Meylan}}]{2012A&A...538A..99S}
{Sluse}, D., {Chantry}, V., {Magain}, P., {Courbin}, F., \& {Meylan}, G. 2012,
  \aap, 538, A99

\bibitem[{{Suyu} {et~al.}(2010){Suyu}, {Marshall}, {Auger}, {Hilbert},
  {Blandford}, {Koopmans}, {Fassnacht}, \& {Treu}}]{2010ApJ...711..201S}
{Suyu}, S.~H., {Marshall}, P.~J., {Auger}, M.~W., {et~al.} 2010, \apj, 711, 201

\bibitem[{{Suyu} {et~al.}(2012){Suyu}, {Treu}, {Blandford}, {Freedman},
  {Hilbert}, {Blake}, {Braatz}, {Courbin}, {Dunkley}, {Greenhill}, {Humphreys},
  {Jha}, {Kirshner}, {Lo}, {Macri}, {Madore}, {Marshall}, {Meylan}, {Mould},
  {Reid}, {Reid}, {Riess}, {Schlegel}, {Scowcroft}, \&
  {Verde}}]{2012arXiv1202.4459S}
{Suyu}, S.~H., {Treu}, T., {Blandford}, R.~D., {et~al.} 2012, ArXiv e-prints,
  arXiv:1202.4459

\bibitem[{{Suyu} {et~al.}(2013{\natexlab{a}}){Suyu}, {Treu}, {Hilbert},
  {Sonnenfeld}, {Auger}, {Blandford}, {Collett}, {Courbin}, {Fassnacht},
  {Koopmans}, {Marshall}, {Meylan}, {Spiniello}, \&
  {Tewes}}]{2013arXiv1306.4732S}
{Suyu}, S.~H., {Treu}, T., {Hilbert}, S., {et~al.} 2013{\natexlab{a}}, ArXiv
  e-prints, arXiv:1306.4732

\bibitem[{{Suyu} {et~al.}(2013{\natexlab{b}}){Suyu}, {Auger}, {Hilbert},
  {Marshall}, {Tewes}, {Treu}, {Fassnacht}, {Koopmans}, {Sluse}, {Blandford},
  {Courbin}, \& {Meylan}}]{2013ApJ...766...70S}
{Suyu}, S.~H., {Auger}, M.~W., {Hilbert}, S., {et~al.} 2013{\natexlab{b}},
  \apj, 766, 70

\bibitem[{{Tavecchio} {et~al.}(2007){Tavecchio}, {Maraschi}, {Wolter},
  {Cheung}, {Sambruna}, \& {Urry}}]{2007ApJ...662..900T}
{Tavecchio}, F., {Maraschi}, L., {Wolter}, A., {et~al.} 2007, \apj, 662, 900

\bibitem[{{Tewes} {et~al.}(2013{\natexlab{a}}){Tewes}, {Courbin}, \&
  {Meylan}}]{2013A&A...553A.120T}
{Tewes}, M., {Courbin}, F., \& {Meylan}, G. 2013{\natexlab{a}}, \aap, 553, A120

\bibitem[{{Tewes} {et~al.}(2013{\natexlab{b}}){Tewes}, {Courbin}, {Meylan},
  {Kochanek}, {Eulaers}, {Cantale}, {Mosquera}, {Magain}, {Van Winckel},
  {Sluse}, {Cataldi}, {V{\"o}r{\"o}s}, \& {Dye}}]{2013A&A...556A..22T}
{Tewes}, M., {Courbin}, F., {Meylan}, G., {et~al.} 2013{\natexlab{b}}, \aap,
  556, A22

\bibitem[{{Tian} {et~al.}(2013){Tian}, {Ko}, \& {Chiu}}]{2013ApJ...770..154T}
{Tian}, Y., {Ko}, C.-M., \& {Chiu}, M.-C. 2013, \apj, 770, 154

\bibitem[{{Tonry}(1991)}]{1991ApJ...373L...1T}
{Tonry}, J.~L. 1991, \apjl, 373, L1

\bibitem[{{Tortora} {et~al.}(2004){Tortora}, {Piedipalumbo}, \&
  {Cardone}}]{2004MNRAS.354..343T}
{Tortora}, C., {Piedipalumbo}, E., \& {Cardone}, V.~F. 2004, \mnras, 354, 343

\bibitem[{{Treu} \& {Koopmans}(2002)}]{2002MNRAS.337L...6T}
{Treu}, T., \& {Koopmans}, L.~V.~E. 2002, \mnras, 337, L6

\bibitem[{{Ulrich} {et~al.}(1997){Ulrich}, {Maraschi}, \&
  {Urry}}]{1997ARA&A..35..445U}
{Ulrich}, M.-H., {Maraschi}, L., \& {Urry}, C.~M. 1997, \araa, 35, 445

\bibitem[{{Vuissoz} {et~al.}(2007{\natexlab{a}}){Vuissoz}, {Courbin}, {Sluse},
  {Meylan}, {Ibrahimov}, {Asfandiyarov}, {Stoops}, {Eigenbrod}, {Le Guillou},
  {van Winckel}, \& {Magain}}]{Vuissoz2007}
{Vuissoz}, C., {Courbin}, F., {Sluse}, D., {et~al.} 2007{\natexlab{a}}, \aap,
  464, 845

\bibitem[{{Vuissoz} {et~al.}(2007{\natexlab{b}}){Vuissoz}, {Courbin}, {Sluse},
  {Meylan}, {Ibrahimov}, {Asfandiyarov}, {Stoops}, {Eigenbrod}, {Le Guillou},
  {van Winckel}, \& {Magain}}]{2007A&A...464..845V}
---. 2007{\natexlab{b}}, \aap, 464, 845

\bibitem[{{Vuissoz} {et~al.}(2008{\natexlab{a}}){Vuissoz}, {Courbin}, {Sluse},
  {Meylan}, {Chantry}, {Eulaers}, {Morgan}, {Eyler}, {Kochanek}, {Coles},
  {Saha}, {Magain}, \& {Falco}}]{Vuissoz2008}
---. 2008{\natexlab{a}}, \aap, 488, 481

\bibitem[{{Vuissoz} {et~al.}(2008{\natexlab{b}}){Vuissoz}, {Courbin}, {Sluse},
  {Meylan}, {Chantry}, {Eulaers}, {Morgan}, {Eyler}, {Kochanek}, {Coles},
  {Saha}, {Magain}, \& {Falco}}]{2008A&A...488..481V}
---. 2008{\natexlab{b}}, \aap, 488, 481

\bibitem[{{Walsh} {et~al.}(1979){Walsh}, {Carswell}, \&
  {Weymann}}]{1979Natur.279..381W}
{Walsh}, D., {Carswell}, R.~F., \& {Weymann}, R.~J. 1979, \nat, 279, 381

\bibitem[{{Winn} {et~al.}(2002){Winn}, {Kochanek}, {McLeod}, {Falco}, {Impey},
  \& {Rix}}]{2002ApJ...575..103W}
{Winn}, J.~N., {Kochanek}, C.~S., {McLeod}, B.~A., {et~al.} 2002, \apj, 575,
  103

\bibitem[{{Wu} {et~al.}(2014){Wu}, {Su}, {Bravar}, {Chang}, {Fan}, {Pohl}, \&
  {Walter}}]{2014arXiv1407.0710W}
{Wu}, X., {Su}, M., {Bravar}, A., {et~al.} 2014, ArXiv e-prints,
  arXiv:1407.0710

\bibitem[{{Wucknitz} {et~al.}(2004){Wucknitz}, {Biggs}, \&
  {Browne}}]{2004MNRAS.349...14W}
{Wucknitz}, O., {Biggs}, A.~D., \& {Browne}, I.~W.~A. 2004, \mnras, 349, 14

\bibitem[{{York} {et~al.}(2005){York}, {Jackson}, {Browne}, {Wucknitz}, \&
  {Skelton}}]{2005MNRAS.357..124Y}
{York}, T., {Jackson}, N., {Browne}, I.~W.~A., {Wucknitz}, O., \& {Skelton},
  J.~E. 2005, \mnras, 357, 124

\end{thebibliography}
\end{document}